\UseRawInputEncoding
\pdfoutput=1
\pdfoutput=1
\pdfoutput=1
\pdfoutput=1
\pdfoutput=1

\documentclass[conference,compsoc]{IEEEtran}

\IEEEoverridecommandlockouts
\usepackage{algorithm} 
\usepackage{algorithmic}
\usepackage{booktabs}
\usepackage{clrscode3e}
\usepackage{hyperref}
\usepackage{multirow}
\usepackage{graphicx}
\usepackage{subfigure}
\usepackage{textcomp}
\usepackage{xcolor}
\usepackage{acronym}
\usepackage{amsmath}

\usepackage{amssymb}
\usepackage{amsfonts}
\usepackage{amsthm}
\usepackage{mathrsfs}
\usepackage{indentfirst}
\usepackage{cases}
\usepackage{bm}
\usepackage{color}
\usepackage{stfloats}
\usepackage{url}
\usepackage{verbatim}
\usepackage{array}
\usepackage{epstopdf}
\acrodef{SR}{Sequential Recommendation}
\acrodef{EA-GCL}{External Attention-enhanced Graph Contrastive Learning}
\acrodef{GCL}{Graph Contrastive Learning}
\acrodef{CDR}{Cross-Domain Recomendation}
\acrodef{CR}{Cross-domain Recommendation}
\acrodef{CSR}{Cross-domain Sequential Recommendation}
\acrodef{GR}{Group Recommendation}
\acrodef{MDP}{Markov Decision Process}
\acrodef{GRU}{Gated Recurrent Unit}
\acrodef{MLP}{Multi-Layer Perceptron}
\acrodef{GNN}{Graph Neural Network}
\acrodef{GCNs}{Graph Convolution Networks}
\acrodef{GCN}{Graph Convolution Network}
\acrodef{SCR}{Single-target Cross-domain Recommendation}
\acrodef{DCR}{Dual-target Cross-domain Recommendation}
\acrodef{RNN}{Recurrent Neural Network}
\acrodef{CNN}{Convolutional Neural Network}
\acrodef{DNN}{Deep Neural Network}
\acrodef{MRR}{Mean Reciprocal Rank}
\acrodef{CF}{Collaborative Filtering}
\acrodef{SA}{Self-Attention}
\acrodef{CDS}{Cross-Domain Sequential}
\acrodef{SSL}{Self-Supervised Learning}
\acrodef{GAN}{Generative Adversarial Network}
\acrodef{BU}{Balance Unit}
\acrodef{EA}{External Attention}

\ifCLASSOPTIONcompsoc
  \usepackage[nocompress]{cite}
\else
  \usepackage{cite}
\fi
\ifCLASSINFOpdf
\else
\fi
\begin{document}
% \title{EA-GCL: External Attention-enhanced Graph Contrastive Learning for Unbiased Cross-domain Sequential Recommendation}
\title{Unbiased and Robust: External Attention-enhanced Graph Contrastive Learning for Cross-domain Sequential Recommendation}

%\author{\IEEEauthorblockN{Anonymous}
%作者信息
\author{
    \IEEEauthorblockN{Xinhua Wang\thanks{*~Co-corresponding authors.}\IEEEauthorrefmark{2}$^*$, Houping Yue\IEEEauthorrefmark{2}, Zizheng Wang\IEEEauthorrefmark{3}, Liancheng Xu\IEEEauthorrefmark{2}, Jinyu Zhang\IEEEauthorrefmark{4}$^*$\footnote{Co-Corresponding authors.}
    }
    \IEEEauthorblockA{
        \IEEEauthorrefmark{2}School of Information Science and Engineering, Shandong Normal University, China
    }
    \IEEEauthorblockA{
        \IEEEauthorrefmark{3}Zhongtai Securities Institute for Financial Studies, Shandong University, China
    }
    \IEEEauthorblockA{
        \IEEEauthorrefmark{4}School of Computer Science and Engineering, Shandong University of Science and Technology, China\\
        wangxinhua@sdnu.edu.cn, yuehouping@gmail.com, JohnWangzz731@gmail.com,\\ lchxu@163.com, jinyuz1996@outlook.com
    }
}

\maketitle

\begin{abstract}
Cross-domain sequential recommenders (CSRs) are gaining considerable research attention as they can capture user sequential preference by leveraging side information from multiple domains. However, these works typically follow an ideal setup, i.e., different domains obey similar data distribution, which ignores the bias brought by asymmetric interaction densities (a.k.a. the inter-domain density bias). Besides, the frequently adopted mechanism (e.g., the self-attention network) in sequence encoder only focuses on the interactions within a local view, which overlooks the global correlations between different training batches. To this end, we propose an External Attention-enhanced Graph Contrastive Learning framework, namely EA-GCL. Specifically, to remove the impact of the inter-domain density bias, an auxiliary Self-Supervised Learning (SSL) task is attached to the traditional graph encoder under a multi-task learning manner. To robustly capture users' behavioral patterns, we develop an external attention-based sequence encoder that contains an MLP-based memory-sharing structure. Unlike the self-attention mechanism, such a structure can effectively alleviate the bias interference from the batch-based training scheme. Extensive experiments on two real-world datasets demonstrate that EA-GCL outperforms several state-of-the-art baselines on CSR tasks. The source codes and relevant datasets are available at https://github.com/HoupingY/EA-GCL.
\end{abstract}


\begin{IEEEkeywords}
Cross-domain sequential recommendation, Unbiased recommender system, Graph neural networks, Attention mechanism, Contrastive learning 
\end{IEEEkeywords}

% \fi
\IEEEpeerreviewmaketitle

\section{Introduction \label{sec:introduction}}
% first paragraph
\noindent \ac{SR} aims to model the evolution of user behaviors from their historical interaction sequences, which has achieved state-of-the-art performance in various practical scenarios (e.g., e-commerce platform~\cite{XiaHXP23}, online retrieval~\cite{ZhangLXXLHCM22}, and television service~\cite{Yang2023unbias}). Nowadays, users prefer to register on different platforms to access domain-specific services~\cite{Guo2022TiDA}. Hence, their historical interactions in multiple domains will be entangled in several hybrid sequences, resulting in poor recommendation performance. To this end, \acf{CSR} methods perform better than Sequential Recommenders (SRs) as they can consider both the cross-domain and sequential characteristics~\cite{Zhang2023Cross, SunMRLCRMR23}. The purpose of \ac{CSR} is to leverage the side information from different domains to recommend the next item for overlapped users (i.e., the users whose historical behaviors can be observed on multiple platforms), thus making it possible for traditional SRs to process sparse data or cold-start scenarios. 

% 第二段
% 1
Recent studies on \ac{CSR} adopt RNN~\cite{SunMRLCRMR23}, GCN~\cite{Zhang2023Cross,ZhengSL022} or Transformer~\cite{Li_dual_KDD_2021,ChenGS21} to model the complicated user-item interactions from hybrid sequences and transfer domain-specific knowledge between different domains. But they are typically following an ideal setup (i.e., they suppose different domains are subject to similar data distribution~\cite{ZhengSL022,LiuSCZ21}). However, the interaction densities in multiple domains are usually asymmetric, probably because different types of interactions require divergent time consumption (e.g., it takes more time to finish reading a book than a movie). In this case, due to the significant difference in interaction density, traditional cross-domain recommenders may be more inclined to consider preferences from the domain with denser interactions while learning representation for users, ignoring the critical preferences from the relatively sparse domain. In this paper, We define such interference as the inter-domain density bias.
Besides, existing \ac{CSR} methods usually exploit the self-attention network to calculate the correlations among items while learning sequence representations for overlapped users~\cite{ChenGS21,Guo2022TiDA}. Since these methods typically follow the batch-based training scheme, the self-attention mechanism in their sequence encoder can only consider the local associations of items within a single batch but ignore the global relevance between different batches, resulting in biased sequence representation learning. Hence, providing unbiased cross-domain sequential recommendations is an urgent yet ongoing trend, but contacts two inevitable challenges: 
1) Existing \ac{CSR} methods generally suffer from the inter-domain density bias, reducing the reliability of cross-domain knowledge transferring. 
2) The frequently used self-attention mechanism in \ac{CSR} has difficulty calculating the item correlations between different training batches, leading to a biased sequence representation learning.

% 第三段
% 1
In this work, we introduce an external attention-enhanced graph contrastive learning framework to address the above challenges named EA-GCL.
Specifically, we construct \ac{CDS} graphs to model user interactions from two domains and the sequential relationships between items. Then, we treat NGCF~\cite{Wang0WFC19} as the basic graph encoder that propagating domain-specific message on the \ac{CDS} graphs. To alleviate the interference of inter-domain density bias, we attach an auxiliary \ac{SSL} task to the graph encoder with a multi-task training strategy. By perturbing the structure of the \ac{CDS} graph in two different ways (i.e., Item Dropout (ID) and Sequence Reorder (SR)), the \acf{GCL}-based framework can generate additional self-supervised signals to enhance the node representation learning.
% Moreover, to simultaneously optimize the traditional node representation learning and the \ac{GCL} auxiliary task, a multi-task training strategy is adopted.
% 6
% As proved in the latest works, the augmentation via perturbing the graph structure may affect the original users' inner-domain preferences~\cite{Yu2021graphCL,Lee0P22}. Hence, \ac{EA-GCL} only chooses to perturb the domain which contains relatively denser interactions to avoid losing significant collaborative filtering signals in the domain with sparser density.
% 8
Besides, inspired by the successful application of \ac{EA}~\cite{GuoLMH23} in Computer Vision (CV), we devise a parallel external attention network to capture users’ sequential patterns from hybrid sequences without worrying about losing the global correlations of items between different training batches. Unlike the traditional self-attention mechanism, it is a simple, external, but efficient attention network with an independent MLP-based memory-sharing structure, which attentively considers items’ correlations between different batches and learns a more robust sequence representation for users.
% 9
% Consequently, by jointly optimizing the graph encoder and the sequence encoder, \ac{EA-GCL} can provide unbiased cross-domain sequential recommendations to overlapped users by considering their hybrid interaction sequences.
The main contributions of this work can be summarized as follows:
\begin{itemize}
\item In this work, we investigate and explicitly define two types of biases that existed in the \acf{CSR} scenario: 1) the inter-domain density bias that affects the cross-domain information sharing; 2) the bias brought by the traditional batch-based training scheme.

\item We propose a novel \acf{EA-GCL} framework to address the above issues. Specifically, EA-GCL eliminates the inter-domain density bias by combining the traditional node representation learning with an auxiliary \ac{SSL} task. Then, we devise an innovative MLP-based \acf{EA} network to calculate the associations among items between different training batches.

\item Extensive experiments have been conducted on two real-world datasets, and the experimental results demonstrate the superiority of \ac{EA-GCL} in providing unbiased \ac{CSR}.

\end{itemize}
\section{Related Work \label{sec:relatedwork}}
\subsection{Sequential Recommendation}
% Sequential Recommendation
% 1
\noindent \acf{SR} aims to capture the evolution of users' preferences from their historical behavioral sequences~\cite{Xie2022sequential}. 
% 2
In prior studies, \acf{RNN}~\cite{hidasi2016srnn,quadrana2017hrnn,Qiu2022RNN} has been widely applied to \ac{SR} tasks as RNNs are capable of processing input of any length. 
% 3
However, these methods meet gradient vanishing problems while processing long sequences. 
% 4
To avoid the above limitation and improve the performances of Sequential Recommenders (SRs), researchers put their efforts into Generative Adversarial Networks (GANs)~\cite{Lv00FL21}, Graph Neural Networks (GNNs) ~\cite{ZhangWYLW23,ZhangLXXLHCM22}, and Transformers~\cite{XiaHXP23}, etc. 
% 5
These models achieve state-of-the-art performances by leveraging deeper and more complicated network structures to learn the long-term preferences of users~\cite{ZhangWYLW23}. 
% 6
Nevertheless, they ignore the data sparsity problems and have difficulty providing recommendations on the cold-start scenario.

\subsection{Cross-domain Sequential Recommendation}
% Cross-domain Sequential Recommendation
% 1
\noindent By further considering the cross-domain characteristics while modeling user sequential preference, \acf{CSR}~\cite{Zhang2023Cross,ZhengSL022,CaoCSLW22} provides an effective solution to the cold-start challenge.
% 2
The core idea of \ac{CSR} is to learn the evolution of user preferences from their hybrid behavioral sequences so as to provide accurate recommendations for overlapped users in multiple domains.
% 3
In early explorations, Ma et al.~\cite{ma2019pi} and Sun et al.~\cite{SunMRLCRMR23} employ a \ac{GRU}-based parallel information-sharing network to capture users' sequential behavior. 
% 4
Then, Guo et al.~\cite{Guo2022TiDA} and Zhang et al.~\cite{Zhang2023Cross} raise the concept of the \acf{CDS} graph and leverage \ac{GCN} to learn cross-domain node representation from the graph.
% 5
Moreover, Li et al.~\cite{Li_dual_KDD_2021} and Chen et al.~\cite{ChenGS21} bring the superiority of attention mechanisms to model users' domain-specific knowledge among their hybrid interaction sequences. 
% 6
These methods have a powerful capability of representation learning but overlook the impact of the biases which existed in the attribute of domains and sequences.

\subsection{Unbiased Recommendation}
% Unbiased Recommendation
% 1
\noindent Recently, several studies have put their efforts to mitigate the impact of bias on the recommender system yet make it more robust to adapt various application scenarios~\cite{PitouraSK21}. 
% 2
In traditional recommendation tasks, a popular research idea is to analyze item attributes and enhance collaborative filtering signals to counteract popularity bias~\cite{Wan00WGT22,XvLL0LH22}. 
% 3
In \acf{SR} tasks, the observed sequences may also be contaminated by exposure or selection biases~\cite{WangSWCCW22}. 
% 4
To address the above issues, researchers employ global contrastive learning~\cite{Yang2023unbias} or inverse propensity score (IPS)~\cite{WangSWCCW22} to alleviate the interference of biases.
% 5
In \acf{CR} scenarios, the exploration of related unbiased recommendation methods is relatively scarce.
% 6
To our knowledge, DL-CR~\cite{Li2021debias} and SCDGN~\cite{LiAZHHYK22} are the only two proposed debiasing works for cross-domain scenarios, which define bias in cross-domain scenarios as differences in domain structure and mitigate its effects with the help of a causality-based IPS estimator and a graph clustering strategy, respectively. 
% 7
Despite the remarkable performance achieved by the above methods for providing unbiased recommendations, the inter-domain density bias, and the bias brought by traditional batch-based training scheme, remain largely unexplored.

\section{Method \label{sec:method}}
\noindent In this section, we first introduce symbol definitions used in this paper and the task definition for~\acf{CSR}. Then, we introduced \ac{EA-GCL} in detail from four aspects: node representation learning, graph contrastive learning, sequence representation learning, and final prediction.
\begin{figure*}[ht]
    \centering
    \includegraphics[width=15cm]{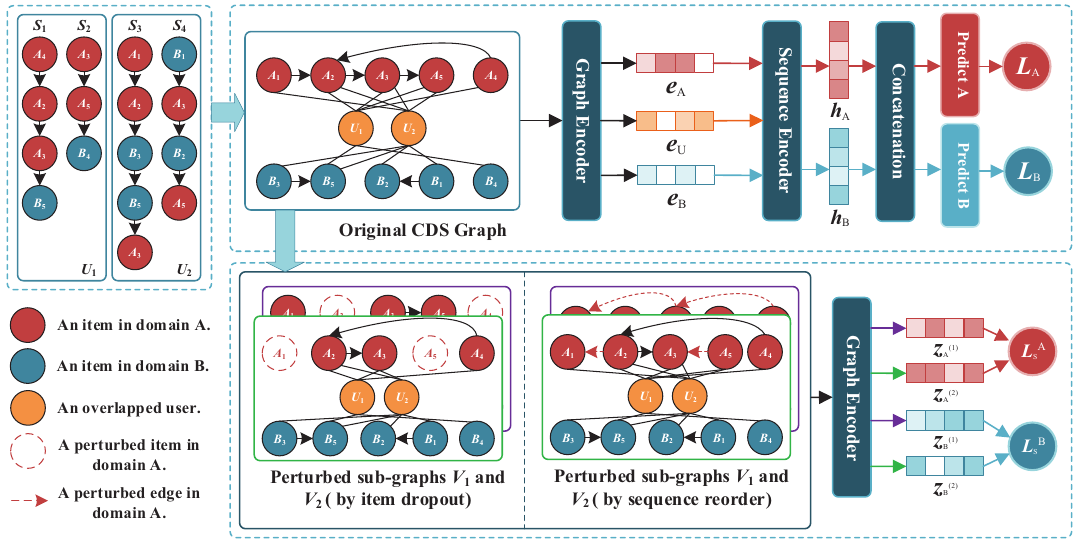} %width=15
    \caption{shows the overview of \ac{EA-GCL}, where $\bm{S}_1$ to $\bm{S}_4$ denote the hybrid interaction sequences of two overlapped users $\bm{U}_1$ and $\bm{U}_2$.}
    \label{fig1:overview}
\end{figure*}
\subsection{Preliminary \label{subsec:preliminary}}
\noindent Suppose $\mathcal{A}$ is a domain with denser user interactions while domain $\mathcal{B}$ contains relatively sparser density. To better formulate functions of \ac{EA-GCL}, we define the original input hybrid sequence for an overlapped user as $S_h$. Then we further extract the sub-sequences from $S_h$ with item orders fixed for domain $\mathcal{A}$ and domain $\mathcal{B}$, which can be denoted as $S_A=\{A_1, A_2, \dots, A_i, \dots, A_m\}$ and $ S_B=\{B_1, B_2, \dots, B_j, \dots, B_n\} $, respectively, where the $A_i \in\mathcal{A}$ $(1\leq i\leq m)$ denotes an item in domain $\mathcal{A}$ and $B_j \in  \mathcal{B}$ $(1\leq j\leq n)$ represents an item in domain $\mathcal{B}$. Let $ U = \{U_1,U_2, \dots, U_k, \dots, U_p\}$ be the set of overlapped users (i.e., the users who have historical behaviors in both domains), where $U_k \in \mathcal{U}$ $(1\leq k\leq p)$ denotes an independent user in $\mathcal{U}$.

The specific definition of \ac{CSR} task is to recommend next items for overlapped users on both domains by modeling their hybrid sequential interactions. The recommendation probabilities for all candidate items in $\mathcal{A}$ and $\mathcal{B}$ can be respectively denoted as:
\begin{align}
& P(A_{i+1}|S_A, S_B)\sim f_A(S_A, S_B), \\
& P(B_{j+1}|S_B, S_A)\sim f_B(S_B, S_A),
\end{align}
where $P(A_{i+1}|S_A, S_B)$ is the probability of recommending $A_{i+1}$ as the next consumed item in domain $\mathcal{A}$ based on $S_A$ and $S_B$. $f_A(S_A,S_B)$ denotes the learning function utilized to estimate the probability. Similar definitions for domain $\mathcal{B}$ can be denoted as $P(B_{j+1}|S_B, S_A)$ and $f_B(S_B, S_A)$.
\subsection{Overview of EA-GCL\label{subsec:overview}}
\noindent The mission of EA-GCL is to provide unbiased and robust cross-domain sequence recommendation for overlapping users. As shown in Fig.~\ref{fig1:overview}, we take the same composition rules provided by TiDA-GCN~\cite{Guo2022TiDA} to construct the original \acf{CDS} graph, which can simultaneously model the user-item interactions and the inner-domain sequential orders extracted from the hybrid sequences. Then, a NGCF~\cite{Wang0WFC19} graph encoder is adopted to learn the node representations (i.e., $e_A$, $e_U$, and $e_B$) from the graphs. To address the inter-domain density bias, we attach an auxiliary graph contrastive learning-based component to the graph encoder. By maximizing the agreement between different augmentation views of the same \ac{CDS} graph, the \ac{SSL} auxiliary task can generate sufficient self-supervised signals to learn unbiased node representations for both domains. 
Analogously, instead of adopting self-attention based sequence encoder, we bring a new take on external attention~\cite{GuoLMH23} to attentively learn users' sequential representations via an external MLP-based memory structure, so as to alleviate the interference brought by the batch-based training scheme. 
As it stores the weighting information through external memory units, such a sequence encoder can calculate the correlations among all the items in each sequence without the limitation of batches. Then, \ac{EA-GCL} facilitates the final predictions via a concatenation. We mutually optimize the parameters and maximize the efficiency of the knowledge transferring via a joint training paradigm.
\subsection{Node Representation Learning\label{subsec:origin}}
\noindent In the early explorations~\cite{guo_DAGCN_2021,Zhang2023Cross}, constructing \acf{CDS} graphs has been proven effective in modeling the complicated associations between users and items (i.e., the user-item interactive relationships from both domains and the sequential relationships between different items). In this work, we also adopt a similar strategy to construct \ac{CDS} graphs as $\mathcal{G}=\{G_1, G_2, \dots, G_t\} $, where $t$ is the number of training batches. In detail, overlapped users and items in both domains are denoted as distinct nodes in the graph, while the complicated relationships between them play the role of edges.
% Inspired by NGCF~\cite{Wang0WFC19} and TiDA-GCN~\cite{Guo2022TiDA}, we construct the \acf{CDS} graphs (i.e., $\mathcal{G}=\{G_1, G_2, \dots, G_t\} $) to model the complicated associations between two domains, where $t$ is the number of training batches. In the \ac{CDS} graph, the interacted items from two domains and overlapped users are presented as nodes, while the interactive user-item associations and the sequential orders between items are denoted as edges.
After the graph construction, we adopt NGCF~\cite{Wang0WFC19} (i.e., a multi-layer graph convolution network with the powerful capability of node representation learning) as the basic graph encoder to propagate embeddings on the \ac{CDS} graph. The data flow on $l$-th layer can be denoted as:
\begin{align}
& \bm{e}^{(l)}=H(\bm{e}^{(l-1)}, \mathcal{G}),
\end{align}
where the $\bm{e}^{(l)}\in \mathbb{R}^{(m+p+n)\times d}$ denotes the message aggregation at the $l$-th layer, $\bm{e}^{(l-1)}$ is that of the previous layer. $H(\cdot)$ denotes the function for neighbor aggregation, which can be further formulated as:
\begin{align}
H(\bm{e}^{(l-1)}, \mathcal{G})=\sigma((\bm{M}+\bm{I}) \bm{e}^{(l-1)} \bm{W}_1 + \nonumber \qquad \\
\bm{M} \bm{e}^{(l-1)}\odot \bm{e}^{(l-1)} \bm{W}_2),
\end{align}
where the $\bm{M}\in \mathbb{R}^{(m+p+n)\times(m+p+n)}$ denotes the Laplacian matrix of the \ac{CDS} graphs $\mathcal{G}$ and the $\bm{I}\in \mathbb{R}^{(m+p+n)\times(m+p+n)}$ is an identity matrix in the same dimension. The detailed definitions of $\bm{M}$ are formulated as:
\begin{align}
& \bm{M}={
\left[ \begin{array}{ccc}
\bm{0} & \bm{R}_{A_i}{}_{U_k} & \bm{0}\\
\bm{R}_{U_k}{}_{A_i}     & \bm{R}_{U_k}{}_{U_k}       & \bm{R}_{U_k}{}_{B_j}\\
\bm{0}      & \bm{R}_{B_j}{}_{U_k} & \bm{0},
\end{array} 
\right ]}, 
\end{align}
where $\bm{R}_{A_i}{}_{U_k}\in\mathbb{R}^{m\times p}$ and $\bm{R}_{B_j}{}_{U_k}\in\mathbb{R}^{n\times p}$ are the weight matrices carrying the weights from user neighbors to the target items in domain $\mathcal{A}$ and domain $\mathcal{B}$, respectively; $\bm{R}_{U_k}{}_{A_i}\in\mathbb{R}^{p\times m}$ and $\bm{R}_{U_k}{}_{B_j}\in\mathbb{R}^{p\times n}$ respectively denote two different weight matrices which contain the weights from items neighbors in both domains to the user nodes; $\bm{R}_{U_k}{}_{U_k}\in\mathbb{R}^{p\times p}$ denotes the weight matrix carrying the weights from user nodes to their neighbors.

Thereafter, the node representation learning of~$\bm{e}_{U}^{(l)}$, $\bm{e}_{A}^{(l)}$, and $\bm{e}_{B}^{(l)}$ on $l$-th layers can be formulated as:
\begin{align}
    \bm{e}_{U}^{(l)} = \sum_{k\in U}(\sum_{i\in S_{A}}\bm{m}^{(l)}_{i\rightarrow k}+\sum_{j\in S_{B}}\bm{m}^{(l)}_{j\rightarrow k}+\bm{m}^{(l)}_{k\rightarrow k});\\
    \bm{e}_{A}^{(l)} = \sum_{i\in S_{A}}(\sum_{k\in U} \bm{m}^{(l)}_{k\rightarrow i}); \qquad \qquad\\
    \bm{e}_{B}^{(l)} = \sum_{j\in S_{B}}(\sum_{k\in U} \bm{m}^{(l)}_{k\rightarrow j}), \qquad \qquad
\end{align}
where $\bm{m}_{i\rightarrow k}$ and $\bm{m}_{j\rightarrow k}$ denote the passing message from the items to the overlapped user, $\bm{m}_{k\rightarrow i}$ and $\bm{m}_{k\rightarrow j}$ respectively denote the message transferred from users to items of domain $\mathcal{A}$ and domain $\mathcal{B}$, and the $\bm{m}_{k\rightarrow k}$ represents the self-connection relations of overlapped users. Then, by concatenation and average pooling, we receive the final node representations $\bm{e}_U$, $\bm{e}_A$ and $\bm{e}_B$ as:
\begin{align}
    \bm{e} = \bm{Avg}(\bm{e}^{(1)}\Vert \dots \Vert \bm{e}^{(s)}),
    % \quad\bm{e}_A = \frac{1}{s}(\bm{e}_{A}^{(1)}\Vert \dots \Vert \bm{e}_{A}^{(s)}),\quad\bm{e}_B = \frac{1}{s}(\bm{e}_{B}^{(1)}\Vert \dots \Vert \bm{e}_{B}^{(s)}),
\end{align}
where $\Vert$ is the concatenation operation and $s$ denotes the number of layers.

\subsection{Graph Contrastive Learning}
\noindent As \acf{GCL} can supplement the corresponding self-supervised signals to align the distribution of interactions between domains, we bring its superiority to alleviate the interference of inter-domain density bias. In this section, we specify the data augmentation strategies on graph structure and introduce the process of graph contrastive learning.
\subsubsection{Data Augmentation on Graph Structure}
\noindent The most common practice is to perturb the structure of the original graphs to generate more contrastive views for the following self-supervised learning~\cite{Xie2022Contrastive}. The key idea of such graph augmentation strategy is to generate a pair of contrastive views by perturbing the structural information on the original graph and to obtain additional self-supervised signals by enlarging the structural consistency of the graph itself and reducing its consistency with the comparative views~\cite{Lee0P22}. However, breaking the graph structure is risky. Because of the existence of the inter-domain density bias, domain $\mathcal{B}$ that with relatively sparser interaction density has more vulnerable structure than domain $\mathcal{A}$, which means the above augmentation strategies may exacerbate the already fragile inner-domain preferences on $\mathcal{B}$. Hence, we brings a special take in \ac{EA-GCL}, which only perturbs the nodes and edges in the relatively stable domain $\mathcal{A}$, as the interactions in domain $\mathcal{A}$ are more expressive and anti-jamming.  

In \ac{EA-GCL}, two types of augmentation operations (i.e., "Item Dropout" and "Sequence Reorder"~\cite{Xie2022Contrastive}) are adopted to generate the augmentation views. Since the Laplacian matrices are used for storing user-item interactions, we construct a pair of masking matrices based on different random interventions to realize the above perturbations. Specifically, we attach the masking matrices to the inputted Laplacian matrix, so as to generate different perturbation views $\bm{V}_1$ and $\bm{V}_2$:
\begin{align}
& \bm{V}_1=(\bm{Q}^{(1)} \odot \bm{M} \mid S_A); \quad \bm{V}_2=(\bm{Q}^{(2)} \odot \bm{M} \mid S_A),
\end{align}
where $\bm{Q}^{(1)}$ and $\bm{Q}^{(2)}$ denote two masking matrices that are applied to the original Laplacian matrix. The masking matrices can also be formulated as:
% \begin{align}
% \bm{Q}^{(1)}={
% \left[ \begin{array}{ccc}
% \bm{0} & \bm{T}^{(1)}_{A_i}{}^{}_{U_k} & \bm{0}\\
% \bm{T}^{(1)}_{U_k}{}^{}_{A_i}     & \bm{0}       & \bm{0}\\
% \bm{0}      & \bm{0} & \bm{0},
% \end{array} 
% \right ]};\nonumber\\
%  \\
% \bm{Q}^{(2)}={
% \left[ \begin{array}{ccc}
% \bm{0} & \bm{T}^{(2)}_{A_i}{}^{}_{U_k} & \bm{0}\\
% \bm{T}^{(2)}_{U_k}{}^{}_{A_i}     & \bm{0}       & \bm{0}\\
% \bm{0}      & \bm{0} & \bm{0}, 
% \end{array} 
% \right ]},\nonumber
% \end{align}
\begin{tiny}
\begin{align}
\bm{Q}^{(1)}={
\left[ \begin{array}{ccc}
\bm{0} & \bm{T}^{(1)}_{A_i}{}^{}_{U_k} & \bm{0}\\
\bm{T}^{(1)}_{U_k}{}^{}_{A_i}     & \bm{0}       & \bm{0}\\
\bm{0}      & \bm{0} & \bm{0},
\end{array} 
\right ]}; \quad
\bm{Q}^{(2)}={
\left[ \begin{array}{ccc}
\bm{0} & \bm{T}^{(2)}_{A_i}{}^{}_{U_k} & \bm{0}\\
\bm{T}^{(2)}_{U_k}{}^{}_{A_i}     & \bm{0}       & \bm{0}\\
\bm{0}      & \bm{0} & \bm{0}, 
\end{array} 
\right ]},\nonumber
\end{align}
\end{tiny}
where $\bm{T}_{A_i}{}_{U_k}$, and $\bm{T}_{U_k}{}_{A_i}$ represent the perturbed weight matrices in two augmentation views. The advantage of using the matrix-form intervention is that we can directly disturb the adjacent matrix, that is, both the nodes and their edges can be perturbed simultaneously. The augmentation strategies we adopted can be further detailed as:

\textbf{Item Dropout (ID).~}
Suppose the length of $S_A$ is $L$, we randomly select $L \times \alpha$ items and then set them to $\bm{0}$ (i.e., the dropout operation), where $\alpha$ is a hyper-parameter that controls the degree of perturbation while generating the augmentation views. In this case, we can add the zero masking into $\bm{T}_{A_i}{}_{U_k}$, and $\bm{T}_{U_k}{}_{A_i}$ on $\bm{Q}^{(1)}$ and $\bm{Q}^{(2)}$.

\textbf{Sequence Reorder (SR).~}
Suppose the length of $S_A$ is $L$, we randomly remove $L \times \alpha$ items from $S_A$ at first and then add them to the end of the $S_A$. In this way, we can disrupt the original item orders in $S_A$ and provide structural disturbance in $\bm{T}_{A_i}{}_{U_k}$, and $\bm{T}_{U_k}{}_{A_i}$.

We fine-tune the $\alpha$ as a hyper-parameter in Section~\ref{subsec:paramanalysis} to further illustrate the impact of node disturbance degree. Thereafter, we learn the node representation on augmentation views by the same basic graph encoder, which can be formulated as:
\begin{align}
& \bm{z}_{1}^{(l)}=H(\bm{z}_{1}^{(l-1)}, \bm{V}_1);\quad\bm{z}_{2}^{(l)}=H(\bm{z}_{2}^{(l-1)}, \bm{V}_2).
\end{align}

Finally, the resulting node representation for items of both domains can be denoted as $\bm{z}_A^{(1)}, \bm{z}_A^{(2)}$ and $\bm{z}_B^{(1)}, \bm{z}_B^{(2)}$, respectively.
In experiments, we introduce \ac{EA-GCL} with "Item Dropout" and "Sequence Reorder" as EA-GCL (ID) and EA-GCL (SR), respectively. Their performance are reported in Table~\ref{tab:resutls} and Table~\ref{tab:ablation}.

\subsubsection{Contrastive Learning}
\noindent After the construction of augmentation views, we treat the views of the same batch (take domain $\mathcal{A}$ as an example) as the positive pairs (i.e., $\{(\bm{z}^{(1)}_{Ao}, \bm{z}^{(2)}_{Ao}) \mid o\in \mathcal{O}\}$), and the views of any different batches as the negative pairs (i.e., $\{(\bm{z}^{(1)}_{Ao}, \bm{z}^{(2)}_{Aq}) \mid o,q\in \mathcal{O},~o\neq q\}$). The auxiliary supervision signals from positive pairs encourage the consistency between different views of the same batch for prediction, while the supervision signals from the negative pairs enforce the divergence among different batches. 
Formally, we adopt the InfoNCE~\cite{He0WXG20} as the contrastive learning loss to maximize the agreement of positive pairs and minimize that of negative pairs:
\begin{align}
    \bm{L}_s^A = \sum_{o\in \mathcal{O}} -~log~ \frac{exp(C(z^{(1)}_{Ao}, z^{(2)}_{Ao})/\tau)}{\sum_{q\in \mathcal{O}} exp(C(z^{(1)}_{Ao}, z^{(2)}_{Aq})/\tau)},
\end{align}
where the $C(\cdot)$ measures the cosine similarity between two vectors; $\tau$ is a regularization parameter of the softmax function. We also adopt a similar definition for $\bm{L}^B_s$ to domain $\mathcal{B}$.

By adopting the multi-task learning strategy, the traditional node representation learning can be enhanced by the auxiliary contrastive learning task that eliminates the inter-domain density bias from the transferred knowledge. We testify its effectiveness in ablation studies at Section~\ref{subsec:ablation}.
\begin{figure}[ht]
    \centering
    \includegraphics[width=8cm]{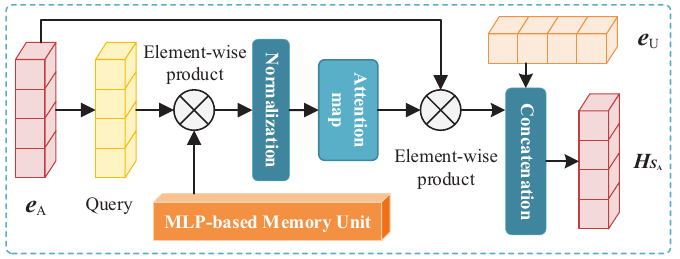}
    \caption{shows the details of the external attention-based sequence encoder (take domain $\mathcal{A}$ as an example here).}
    \label{fig4:seq_encoder}
\end{figure}
\subsection{Sequence Representation Learning}
\noindent In \acf{SR} tasks, self-attention mechanism is frequently exploited to model users' sequential behavioral patterns~\cite{SunLFWZ23,XiaHXP23}, as it uses the combination of self values to refine users' sequence-level representations. However, such weighting strategy only considers the relations between items within one sequence but ignores implied correlations between different batches~\cite{GuoLMH23}, which leads to a biased representation learning.

% However, the learned item representations for both domains still contain the inner-domain popularity bias. 
To this end, we devise an external attention mechanism with MLP-based memory units, which can attentively measure the correlations between items for each sequence yet with no interference of the batch-based training scheme. Taking domain $\mathcal{A}$ as an example, we first treat the item embedding $e_A$ itself as a query vector, then we feed it to the external MLP-based memory-sharing network. The consistency between $A_i$ with $A_j$ ($i, j\in S_A,~i \neq j$) will be calculated in there, which can be further formulated as:
\begin{align}
    a_{i,j} = Norm(f({\bm{e}_{A_i}, \bm{e}_{A_j}})) \qquad \qquad \\
    f(\bm{e}_{A_i}, \bm{e}_{A_j})=\bm{W_2}^T LeakReLU( \qquad \qquad \nonumber\\
    \bm{W_1} (\bm{e}_{A_i}\odot \bm{e}_{A_j}) + \bm{b}),
\end{align}
where the $f(\cdot)$ is an attention function, $\bm{W_1}$ and $\bm{W_2}$ are two weighting matrices in MLP network that maps the inputs into a latent space, $\bm{b}$ is the bias vector. Then, the sequence-level representation for items $\bm{h}_{S_A}$ can be detailed as:
\begin{align}
    a_{i,j} = \frac{exp(f(\bm{e}_{A_i}, \bm{e}_{A_j}))}{\sum_{j=1}^{S_A} \sqrt{exp(f(\bm{e}_{A_i}, \bm{e}_{A_j}))}}; \\
    \bm{h}_{S_A} = \sum_{i=1}^{S_A} a_{i} \bm{e}_i. \qquad \quad
\end{align}
Whenever a new batch of data is input, it interacts with the external MLP network and calculates the correlations between items. Since the calculation of attention score is carried out and stored in external units, each calculation can be considered as a global weighting process, avoiding the bias caused by batch-based training strategies.

After concatenating the embeddings of overlapped users and items, we result in the final sequence-level preference $\bm{H}_{S_A}$:
\begin{align}
    & \bm{H}_{S_A} = Concat[\bm{h}_{S_A},\bm{e}_U]^T,
\end{align}
and we also take the same strategy for domain $\mathcal{B}$ to learn the $\bm{H}_{S_B}$. 

\subsection{Prediction Layers \& Loss Functions}
\noindent After the sequence-level representation learning, we are capable to feed $\bm{H}_{S_A}$ and $\bm{H}_{S_B}$ into the prediction layer as:
\begin{align}
P(A_{i+1}|S_A, S_B) = softmax(\bm{W}_A \cdot [\bm{H}_{S_A},\bm{H}_{S_B}]^\mathrm{T}+\bm{b}_A);\\
P(B_{j+1}|S_B, S_A) = softmax(\bm{W}_B \cdot [\bm{H}_{S_B},\bm{H}_{S_A}]^\mathrm{T}+\bm{b}_B),
\end{align}
where $\bm{W}_A$ and $\bm{b}_A$ are the weight matrix of all items in domain $\mathcal{A}$ and the bias term for domain $\mathcal{A}$, respectively; $\bm{W}_B$ and $\bm{b}_B$ are with a familiar definition on domain $\mathcal{B}$. Then we adopt the cross-entropy as the  loss function for the supervised task to train the framework in both domains:
\begin{align}
\bm{L}_A = -\frac{1}{|\mathcal{S}|}\sum_{S_A, S_B \in \mathcal{S}}\sum_{A_i \in S_A}\text{log} P(A_{i+1}|S_A,S_B), \\
\bm{L}_B = -\frac{1}{|\mathcal{S}|}\sum_{S_B, S_A \in \mathcal{S}}\sum_{B_j \in S_B}\text{log}P(B_{j+1}|S_B,S_A),
\end{align}
where $\mathcal{S}$ denotes the training sequences in both domains. To improve the \ac{CSR} with the \ac{SSL} auxiliary task, we leverage a multi-task training strategy on both domains to jointly optimize the supervised loss and the self-supervised loss $\bm{L}_s^A, \bm{L}_s^B$, the joint learning scheme $\bm{L}_{j}$ can be denoted as:
\begin{align}
    \bm{L}_{j}(\theta)= (\bm{L}_A(\theta) + \beta \bm{L}_{s}^A) + (\bm{L}_B(\theta) + \beta \bm{L}_{s}^B),
\end{align}
where $\beta$ denotes the hyper-parameter that controls the participation of the \ac{SSL} task, $\theta$ represents the trainable parameters. All the parameters are learned via the gradient back-propagation algorithm in an end-to-end training manner.

\section{Experiment \label{sec:experiment}}
\noindent 
% We have conducted a series of experiments on two real-world datasets to verify the effectiveness of \ac{EA-GCL} and
In this section, we are about to answer the following \textbf{R}esearch \textbf{Q}uestions:
\begin{itemize}
    \item[\textbf{RQ1:~}]How does our proposed graph-based approach perform on~\ac{CSR} tasks compared with other state-of-the-art baseline methods? 
    % Does it work on resisting the density bias for \ac{FCSR} task? 
    \item[\textbf{RQ2:~}]Does our proposed graph contrastive learning work on removing inter-domain density bias? If does, which graph augmentation strategy performs better? Is it helpful to leverage the MLP-based external attention mechanism for unbiased sequence-level representation learning?
\begin{table}[ht]
    \centering
    \footnotesize
     \caption{Statistics of two real-world datasets.}
    \begin{tabular}{l|c|c|c|c}
    \toprule
    \multicolumn{1}{c|}{\textbf{Dataset}}&\multicolumn{2}{c|}{\textbf{DOUBAN}}&       \multicolumn{2}{c}{\textbf{AMAZON}} \\
    \cmidrule{1-5}
    \multicolumn{1}{c|}{\textbf{Domain}}
    & A & B
    & A & B \\
    \midrule
    \#Items &14,636 &2,940 &126,526 &61,362\\
    \#Interactions &607,523 &360,798 &1,678,006 &978,226\\
    \#Density &0.0240 &0.0048 &0.0754 &0.0627\\
    \midrule
    \#Overlapped-users & \multicolumn{2}{c|}{6,582}
    & \multicolumn{2}{c}{9,204} \\
    \#Training-sequence & \multicolumn{2}{c|}{42,062}
    & \multicolumn{2}{c}{90,574} \\
    \#Testing-sequence & \multicolumn{2}{c|}{10,431}
    & \multicolumn{2}{c}{14,463} \\
    \bottomrule
    \end{tabular}
    \label{tab:dataset_statistics}
\end{table}
    \item[\textbf{RQ3:~}]How do the hyper-parameters $\alpha$ and $\beta$ affect the performance of the \ac{SSL} auxiliary task in \ac{EA-GCL}?
    \item[\textbf{RQ4:~}]How is the training efficiency of our graph-based method? Is it scalable when processing large-scale data?
\end{itemize}

\subsection{Experimental setup \label{subsec:experimentalset}}
\subsubsection{Datasets\label{subsub:datasets}}
\noindent We evaluate \ac{EA-GCL} on two real-world datasets (DOUBAN~\cite{Zhang2023Cross} and AMAZON~\cite{Guo2022TiDA}), which are released for \ac{CSR} tasks. For both datasets, we randomly select 80$\%$ of all the hybrid sequences as the training set and the rest 20$\%$ as the testing set. Table~\ref{tab:dataset_statistics} shows the statistics about these two datasets. DOUBAN is a cross-domain dataset which has a huge gap in the interaction density between two domains that can be used to verify the effectiveness of removing the inter-domain density biases. And AMAZON is another cross-domain dataset but only with a little gap in the interaction density between domains, which can be used to test the model performance on regular \ac{CSR} scenarios.

\subsubsection{Evaluation Metrics}
\noindent For the model evaluation, we evaluate each testing instance with three frequently used metrics: RC@10~\cite{Zhang2023Cross}, MRR@10~\cite{Guo2022TiDA}, and NDCG@10~\cite{Zhang2023Cross}. 
Then, we evaluate each test instance with all the metrics and report their average values on Tables.
\begin{table*}[ht]
  \centering
  \normalsize
   \caption{Experiment results (\%) of compared methods on DOUBAN and AMAZON. To verify the improvements of EA-GCL, we conduct paired samples t-tests on each metric and mark the significant improvements ($p< .05$) with $^\dag$.}
   \resizebox{\linewidth}{!}{ 
    \begin{tabular}{l|ccc|ccc|ccc|ccc}
    \toprule
    \midrule
    \multicolumn{1}{c|}{\textbf{Dataset}} & \multicolumn{6}{c|}{\textbf{DOUBAN}} &
    \multicolumn{6}{c}{\textbf{AMAZON}} \\
    % \midrule
    \cmidrule{1-13}
    \multicolumn{1}{c|}{\textbf{Domain}}& \multicolumn{3}{c|}{\textbf{A}} & \multicolumn{3}{c|}{\textbf{B}} & \multicolumn{3}{c|}{\textbf{A}} & \multicolumn{3}{c}{\textbf{B}} \\
    \cmidrule{1-13}
    \multicolumn{1}{c|}{\textbf{Metric (@10)}}& \multicolumn{1}{c}{\textbf{RC}} & \multicolumn{1}{c}{\textbf{MRR}} & \multicolumn{1}{c|}{\textbf{NDCG}}
          & \multicolumn{1}{c}{\textbf{RC}} & \multicolumn{1}{c}{\textbf{MRR}}
          & \multicolumn{1}{c|}{\textbf{NDCG}}&
          \multicolumn{1}{c}{\textbf{RC}} & \multicolumn{1}{c}{\textbf{MRR}}& \multicolumn{1}{c|}{\textbf{NDCG}}
          & \multicolumn{1}{c}{\textbf{RC}} & \multicolumn{1}{c}{\textbf{MRR}} & \multicolumn{1}{c}{\textbf{NDCG}}
          \\
    \midrule
    NCF &69.75 &58.05 &40.26 &35.24 &23.29 &18.28 
    &15.59 &11.30 &6.12 &17.38 &13.20 &5.29\\
    NGCF &79.21&77.82&49.12&67.37&54.41&37.26
    &21.52&18.74&12.08&26.55&25.37&17.45\\
    LightGCN &78.52 &75.35 &48.93 &68.32 &52.11 &37.79 
    &21.67 &17.41 &10.13 &26.49 &24.23 &15.09\\
    \midrule
    GRU4REC &80.13& 75.41 &47.81 &66.66 &54.10 &38.03
    &21.51 &17.11 &9.66 &24.51 &22.13 &12.94\\
    HRNN &81.25 &77.90 &48.88 &68.32 &54.99 &38.93 
    &21.92 &17.30 &9.87 &25.10 &22.48 &13.44\\
    NAIS &80.79& 76.87 &49.12 &67.50 &54.14 &39.61
    &22.33 &19.12 &12.99 &24.93 &22.67 &15.94\\
    \midrule
    % CoNet &71.31 &72.92 &40.02&45.53 &37.11 &22.52
    % &18.22 &14.31 &7.80 &19.93 &16.20 &7.69\\
    SGL &82.92 &80.61 &50.18 &72.14 &57.75 & 39.39
    &24.73 &20.31 &14.11 &32.09 &29.66 &19.02\\
    DL-CR &81.85&79.92&50.85&69.33&55.88&38.81
    &24.41 &20.61 &12.88 &28.46 &25.81 &18.41\\
    \midrule
    $\pi$-net &83.22&80.71&51.22&69.54&55.72&39.18
    &24.33 &20.52 &11.80 &27.66 &25.03 &16.20\\
    DA-GCN &83.55 &80.84 &51.53 &71.91 &58.31 &40.67
    &24.62 &20.91 &13.18 &31.12 &28.21 &18.85\\
    TiDA-GCN &83.68 &81.27 &52.02 &72.56 &60.27 &41.38
    &25.05 &21.23 &14.68&32.84 &29.65 &19.12\\
    \midrule
    \textbf{EA-GCL (SR)} &83.69 &81.21 &52.03
    &${75.46}^\dag$ &${67.70}^\dag$ &${44.48}^\dag$
    &{25.03} &${21.97}^\dag$ &${15.11}^\dag$
    &${33.92}^\dag$ &${30.43}^\dag$ &${20.22}^\dag$\\
    \textbf{EA-GCL (ID)} &$\textbf{83.87}^\dag$ &$\textbf{81.43}^\dag$ &$\textbf{52.27}^\dag$
    &$\textbf{76.21}^\dag$ &$\textbf{68.05}^\dag$ &$\textbf{45.60}^\dag$
    &$\textbf{25.38}^\dag$ &$\textbf{22.77}^\dag$ &$\textbf{15.27}^\dag$
    &$\textbf{34.28}^\dag$ &$\textbf{31.74}^\dag$ &$\textbf{20.90}^\dag$\\
    \midrule
    \bottomrule
    \end{tabular}
    }
  \label{tab:resutls}
\end{table*}
\subsubsection{Baselines\label{subsub:baselines}}
\noindent We compare \ac{EA-GCL} with the following baselines from four categories: \textbf{1) Traditional recommendations:~}NCF~\cite{he2017neural}, NGCF~\cite{Wang0WFC19}, LightGCN~\cite{He2020lightGCN}.
% We select several cutting-edge techniques from the current research landscape. (i.e.,NCF~\cite{he2017neural}, NGCF~\cite{Wang0WFC19}, LightGCN~\cite{He2020lightGCN}). These methods are proposed to address the traditional recommendation task, without considering sharing information between domains or learning sequential patterns from data.
%对此句进行删除
%######################################################################################
%We just simply report their recommendation performance on both domains.
%######################################################################################
\textbf{2) Sequential Recommendations:~} GRU4REC~\cite{hidasi2016srnn}, HRNN~\cite{quadrana2017hrnn} and NAIS~\cite{he2018nais}. 
% As these methods are proposed for original sequential recommendations, we intentionally report their performance on two domains to make them comparative.
\textbf{3) Debiasing Recommendations:~}SGL~\cite{WuWF0CLX21} and DL-CR~\cite{Li2021debias}.
% To make these methods comparative, we feed them with sequential inputs and adapt them to the dual-target setting by transferring knowledge between two domains. 
\textbf{4) Cross-domain Sequential Recommendations:~}$\pi$-net~\cite{ma2019pi}, DA-GCN~\cite{guo_DAGCN_2021} and TiDA-GCN~\cite{Guo2022TiDA}. 
\begin{table*}[b]
  \centering
  \caption{Ablation studies on DOUBAN and AMAZON.}
  \footnotesize
  \resizebox{\linewidth}{!}{ 
  \begin{tabular}{l|ccc|ccc|ccc|ccc}
    \toprule
    \midrule
    \multicolumn{1}{c|}{\textbf{Dataset}} & \multicolumn{6}{c|}{\textbf{DOUBAN}} & \multicolumn{6}{c}{\textbf{AMAZON}}\\
    \midrule
    \multicolumn{1}{c|}{\textbf{Domain}} & \multicolumn{3}{c|}{\textbf{A (\%)}} & \multicolumn{3}{c|}{\textbf{B (\%)}} & \multicolumn{3}{c|}{\textbf{A (\%)}} & \multicolumn{3}{c}{\textbf{B (\%)}}\\
    \midrule
    \multicolumn{1}{c|}{\textbf{Metric (@10)}}& \multicolumn{1}{c}{\textbf{RC}} & \multicolumn{1}{c}{\textbf{MRR}} 
    &\multicolumn{1}{c|}{\textbf{NDCG}}& \multicolumn{1}{c}{\textbf{RC}} & \multicolumn{1}{c}{\textbf{MRR}} 
    &\multicolumn{1}{c|}{\textbf{NDCG}}& \multicolumn{1}{c}{\textbf{RC}} & \multicolumn{1}{c}{\textbf{MRR}} 
    &\multicolumn{1}{c|}{\textbf{NDCG}}& \multicolumn{1}{c}{\textbf{RC}} & \multicolumn{1}{c}{\textbf{MRR}} 
    &\multicolumn{1}{c}{\textbf{NDCG}}\\
    \midrule
    GCL (SR)-EA &83.26 &80.11 &49.90 &74.40 &65.59 &42.71 
    &23.99 &20.74 &13.52 &30.56 &28.61 &19.30 \\
    GCL (ID)-EA &83.34 &80.27 &50.08 &75.82 &67.95 &42.75 
    &24.67 &21.02 &13.66 &30.79 &28.80 &19.45 \\
    GCL-CL &83.25 &80.03 &50.74 &72.31 &62.27 &43.91 
    &23.01 &20.54 &13.87 &30.18 &28.15 &19.96 \\
    GCL-ALL &81.18 &78.83 &49.82 &68.60 &56.32 &38.38 
    &22.03 &19.02 &13.11 &27.42 &26.09 &18.96 \\
    \midrule
    \textbf{EA-GCL (SR)} 
    &83.69 &81.21 &52.03 &75.46 &67.70 &44.48
    &25.03 &21.97 &15.11 &33.92 &30.43 &20.22 \\
    \textbf{EA-GCL (ID)} 
    &\textbf{83.87} &\textbf{81.43} &\textbf{52.27} &\textbf{76.21} &\textbf{68.05} &\textbf{45.60} &\textbf{25.38} &\textbf{22.77} &\textbf{15.27} &\textbf{34.28} &\textbf{31.74} &\textbf{20.90} \\
    \midrule
    \bottomrule
    \end{tabular}
    }
  \label{tab:ablation}
\end{table*}
\subsubsection{Implementation Details\label{subsub:implementation}}
% Hard-ware for implementation on EA-GCL.
\noindent \ac{EA-GCL} is implemented by TensorFlow and accelerated by NVIDIA Tesla K80M GPU. We exploit the Xavier~\cite{glorot2010xavier} to initialize model parameters and take Adam~\cite{kingma2014adam} as our optimizing method for the multi-task loss function. 
% Parameters setting for EA-GCL.
For the training settings, we set the batch-size as 256, the learning rate as 0.005, and the dropout rate as 0.1. 
For the basic graph encoder in \ac{EA-GCL}, we both set the embedding-size and the layer-size as 16.
For the \ac{SSL} auxiliary task, we set the regularized parameter ssl-reg as 1e-3. 
% The hyper-parameter $\alpha$ is searched in [0-1] with a step size of 0.1 to adjust the degree of perturbation and we also explore the impact of $\beta\in[0.0,~1.0]$, which controls the participation of the \ac{SSL} loss.
% To adapt the baseline methods, we uniformly change their embedding size to 16. Then, 
We refer to the best hyper-parameter settings reported in each baseline's paper and also fine-tune them on both datasets.

\subsection{Overall Performance (RQ1) \label{subsec:overallperform}}
\noindent Table~\ref{tab:resutls} shows the experimental results of \ac{EA-GCL} over several state-of-the-art baselines on both datasets. And the observations can be summarized as: 
1) Both the \ac{EA-GCL}~(SR) and \ac{EA-GCL}~(ID) outperform all the other baselines on both datasets, demonstrating the superiority of our graph-based solutions in transferring cross-domain knowledge and modeling users' sequential representations.
%缩写后的内容############################################################################
Moreover, \ac{EA-GCL}~(ID) outperforms other methods in both domains, showcasing the effectiveness of our enhanced graph contrastive learning framework in enhancing recommendation performance across dual-target cross-domain settings.
%对下面扩展内容进行缩写####################################################################
%Moreover, \ac{EA-GCL}~(ID) achieves the best performance on both domains, demonstrating that our external-attention enhanced graph contrastive learning framework can simultaneously improve the recommendation performance on both domains under the dual-target cross-domain settings.
%######################################################################################
2) Compared with the most competitive baselines (i.e., the TiDA-GCN), both variants of \ac{EA-GCL} achieve more significant improvements in the domain with relatively sparser densities (i.e., domain B of both datasets), demonstrating the effectiveness of our graph contrastive learning-based solution on alleviating the inter-domain density bias. 
% Furthermore, it also proves that \ac{EA-GCL} can work well in both the biased \ac{CSR} scenario (i.e., the DOUBAN dataset) and the regular \ac{CSR} scenario (i.e., the AMAZON dataset).
% 3) The \ac{CSR} approaches (i.e., $\pi$-net, DA-GCN, and TiDA-GCN) outperform traditional DNN-based recommendation methods (i.e., NCF, NGCF, and LightGCN) and traditional sequential recommendation methods (i.e., GRU4REC, HRNN, and NAIS), indicating the importance of simultaneously considering the sequential and cross-domain characteristics. 
3) \ac{EA-GCL} performs better than all \ac{CSR} baselines, proving the existence of representation biases in \ac{CSR} scenarios and demonstrating the superiority of our unbiased learning paradigm for \ac{CSR} tasks.
% The unbiased recommendation solutions (i.e., SGL and DL-SR) perform better than those traditional methods, which proves that bias does greatly affect the accuracy of recommendations. Besides, 
4)\ac{EA-GCL} outperforms DL-CR, demonstrating that, compared with the cross-domain interaction attribute bias, the inter-domain interaction density bias is more fatal and urgent to be solved.
5) The attention-based methods (i.e., NAIS, TiDA-GCN, and \ac{EA-GCL}) provide a significant improvement on NDCG@10, which indicates that the attention mechanism improve the accuracy of the recall of top items. This also proves that the external attention module in \ac{EA-GCL} does not lose its critical correlation computing ability while obtaining a global perspective. 
\subsection{Ablation Study (RQ2)
\label{subsec:ablation}}
\noindent To further explore the impact of different components to \ac{EA-GCL}, we conduct a series of ablation studies on both datasets with fixed hyper-parameters. The experimental results on DOUBAN and AMAZON are detailed in Table~\ref{tab:ablation}. Specifically, there are four different variants of \ac{EA-GCL} to be introduced:~GCL (SR)-EA and GCL (ID)-EA are two variants of EA-GCL with different graph augmentation strategies (i.e., "Sequence Reorder" and "Item Dropout") and further disable the external attention components; GCL-CL is a variant that removes the auxiliary \acf{SSL} task; GCL-ALL is another variant method that simultaneously disables the graph contrastive learning and the \ac{EA}-based sequence encoder.
From Table~\ref{tab:ablation}, we can observe that: 
1) Both \ac{EA-GCL}~(SR) and\ac{EA-GCL}~(ID) outperform GCL-CL and GCN-ALL, demonstrating the importance of leveraging the \ac{SSL} auxiliary task under the multi-task learning scheme and the significant improvements on both domains (especially on domain B) prove the effectiveness of our GCL-based methods on eliminating the inter-domain density bias. 
2) We notice that \ac{EA-GCL}~(ID) and GCL~(ID)-EA respectively perform better than \ac{EA-GCL}~(SR) and GCL~(SR)-EA, which demonstrates the "Item Dropout" may be a better choice for alleviating inter-domain density bias as it performs more destructive in perturbation operations. 
3) \ac{EA-GCL}~(ID) and \ac{EA-GCL}~(SR) outperform GCL~(ID)-EA and GCL~(SR)-EA on both domains, demonstrating that the global weighting strategy via an external memory structure is effective to resist the bias interference.

\begin{figure*}
    \centering
    \includegraphics[width=16cm]{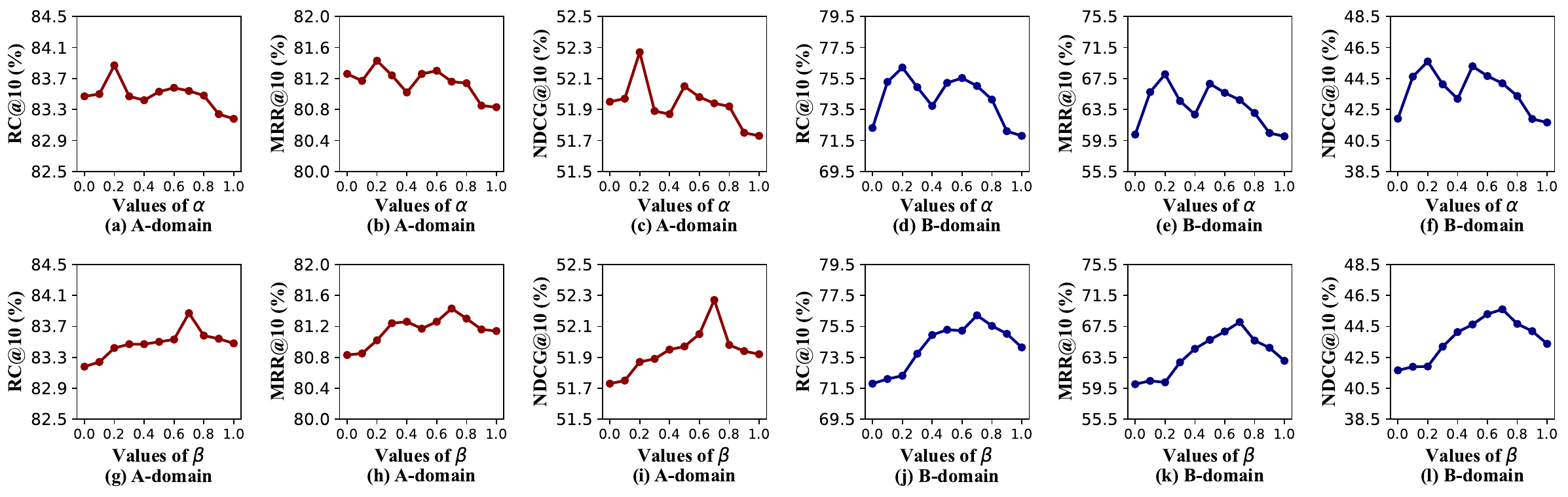} %Fig2_Hyper_params.eps
    \caption{Impact of hyper-parameter $\alpha$ and $\beta$ (take DOUBAN dataset as an example).}
    \label{fig2:hyper_param}
\end{figure*}
\subsection{Additional Analysis \label{subsec:paramanalysis}}
\noindent \textbf{1) Impact of $\bm{\alpha}$ \& $\bm{\beta}$ (RQ3):}~As introduced in Section~\ref{sec:method}, we further illustrate the impact of two hyper-parameters to \ac{EA-GCL} (take \ac{EA-GCL}~(ID) as an example). Fig.~\ref{fig2:hyper_param} shows the performance of \ac{EA-GCL} with different $\alpha\in[0.0,~1.0]$ and $\beta\in[0.0,~1.0]$. Both hyper-parameters exhibit similar trends in all their sub-graphs, but in fact, there is a more significant difference in the sub-graphs of domain B, demonstrating that self-supervised signals from graph contrastive learning have a more dramatic effect on domains with sparser interaction density (i.e., the domain B). 

\begin{figure}[ht]
    \centering
    \includegraphics[width=7cm]{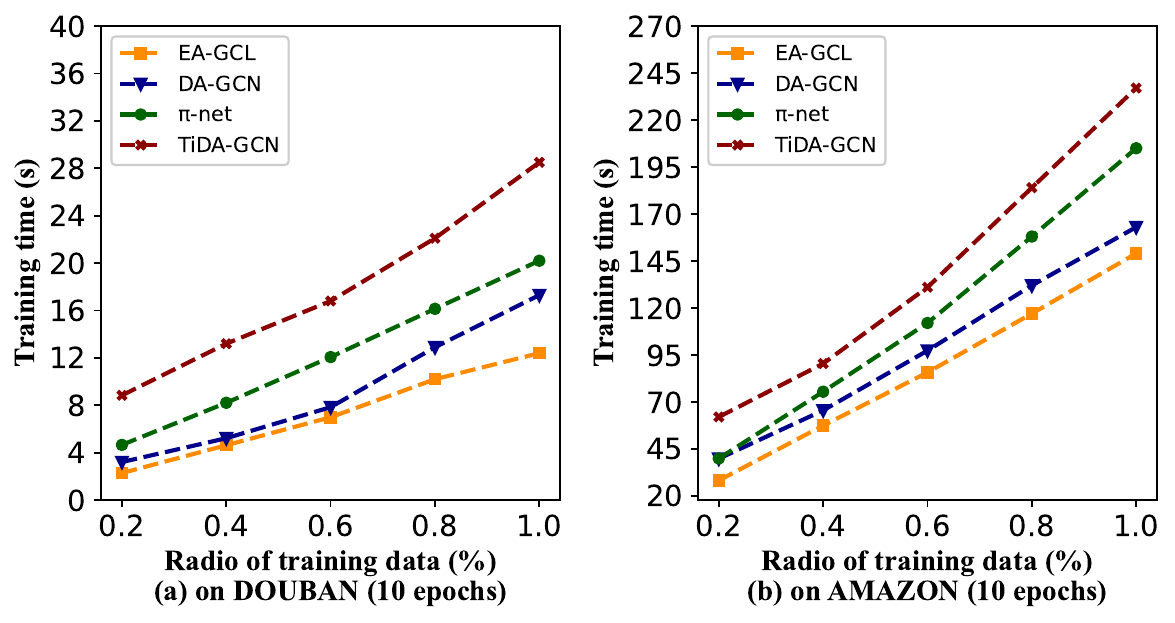}
    \caption{Model Training Efficiency on DOUBAN and AMAZON.}
    \label{fig3:time_cost}
\end{figure}

\noindent \textbf{2) Training Efficiency (RQ4):}
To investigate the training efficiency of \ac{EA-GCL} and testify its scalability on processing large-scale data, we further conduct a series of experiments on DOUBAN and AMAZON to measure the time cost of the model training by changing the proportions of datasets (i.e., ${0.2, 0.4, 0.6, 0.8, 1.0}$). 
% We compare its performance with all the other state-of-the-art \ac{CSR} baselines, the comparison results are reported in 
As shown in Fig.~\ref{fig3:time_cost}, \ac{EA-GCL} costs lower training time than other state-of-the-art \ac{CSR} methods and its time consumption is almost linearly associated with the training ratios, which provides a positive answer to RQ4.
%删除以下内容############################################################################
%As far as we know, graph contrastive learning should have spent more training time due to its auxiliary \ac{SSL} tasks.
%######################################################################################
% However, instead of the traditional attention mechanism, our proposed \ac{EA-GCL} exploits the novel external attention to calculate the correlations between items, which greatly reduces the time consumption during parameter updates. 
%删除以下内容############################################################################
%This may be the reason why \ac{EA-GCL} cost less training time than DA-GCN and TiDA-GCN.
%######################################################################################
% 2) Besides, we also observe that the time consumption of \ac{EA-GCL} is almost linearly associated with the training ratios. It means that the performance of EA-GCL is relatively stable while changing the scale of data, which provides a positive answer to RQ4.

\section{Conclusions\label{sec:conclusion}}
\noindent 
For the first time in \acf{CSR}, this work has investigated both the inter-domain density bias brought by the interactive difference between domains and the training bias brought by the frequently used batch-based training scheme.
To provide unbiased and robust cross-domain sequential recommendations, we introduce an \acf{EA}-enhanced \acf{GCL}-based framework, namely EA-GCL. Specifically, an \ac{SSL} task is attached to traditional graph encoder for alleviating the impact of the inter-domain density bias, which contains two types of graph augmentation methods (i.e., the node dropout and the sequence reorder) that generate auxiliary self-supervised signals to align the data distributions between domains. To robustly learn sequence-level representations under the batch-based training scheme, we devise a novel external attention network to replace the traditional self-attention mechanism. It can attentively capture users' sequential patterns via an MLP-based memory-sharing structure. 
We conduct extensive experiments on two large-scale real-world datasets. After thorough analysis, the experimental results validate the effectiveness of each individual component in EA-GCL and show the superiority of our graph-based approach on \ac{CSR} tasks compared with several state-of-the-art baselines.



%\bibliographystyle{IEEEtran}


% \end{thebibliography}

\bibliographystyle{IEEEtran}
\bibliography{reference}

\end{document}